\documentclass[a4paper,11pt]{article}
\usepackage[utf8x]{inputenc}
\usepackage{amssymb}
\usepackage{amsfonts}
\usepackage{amsmath}
\usepackage{graphicx}
\usepackage{color}

\definecolor{DarkBlue}{rgb}{0,0.1,0.7}

\title{Two-dimensional Dirac fermion in presence of an asymmetric vector potential }
\author{
	A. Ishkhanyan${}^{a,b}$, V. Jakubsk\'y${}^c$\\
{\small \textit{${}^a$Russian-Armenian University, Yerevan, 0051 Armenia}}\\
{\small \textit{${}^b$Institute of Physics and Technology, National Research Tomsk Polytechnic University,}}\\ \small \textit{Tomsk, 634050 Russian Federation }\\
{\small \textit{${}^c$Nuclear Physics Institute of the CAS, \v Re\v z, 25068, Czech Republic}}\\
{\small \textit{E-mails: aishkhanyan@gmail.com, jakub@ujf.cas.cz
} }}

\begin{document}

\maketitle

\begin{abstract}
We introduce the new, exactly solvable model of the two-dimensional Dirac fermion in presence of an asymmetric, P\"oschl-Teller-like vector potential. Utilizing the translation invariance of the system, the effective one-dimensional stationary equation is brought into the form of the Heun equation and its fundamental solutions are found as an irreducible combination of two Gauss hypergeometric functions. The energy spectrum and the scattering is studied in dependence on the conserved longitudinal momentum as well as on the strength of the coupling.
\end{abstract}

\section{Introduction}
Experimental isolation of graphene \cite{grapheneseminal} ignited intensive study of this new material. Many of its remarkable properties, e.g. unconventional quantum Hall effect, minimal conductivity, or light absorption emerge due to the fact that the low-energy excitations of electrons in the lattice behave like mass-less, two-dimensional Dirac fermions \cite{unconventionalQHE}, \cite{GusyninQHE}, \cite{RoomTQHE}, \cite{Katsnelson1}, \cite{KatsnelsonBook}. In the recent years, there have been discovered other materials that share this property, hinted by existence of Dirac cones in their band structure. Let us mention silicene, germanene, dichalcogenides \cite{SiGeTheor1}, \cite{SiGeExp2}, \cite{dichalcogenides}, or artificially prepared materials where the low-energy quasi-particles are described by two-dimensional Dirac equation, see \cite{agr1} for review. These systems were called Dirac materials or Dirac matter in the literature \cite{Wehling}, \cite{DiracMatter}.

Unlike standard semi-conductors, electron transport in Dirac materials cannot be controlled very well by an electrostatic barrier; the charge carriers can tunnel it without backscattering. This phenomenon was understood as the manifestation of Klein tunneling \cite{kleintuneling}. Alternative ways to confinement and control of the charge carriers were proposed. The magnetic field proved to be one of the feasible options \cite{DeMartino}.

Symmetries of the stationary equation can be used for separation of variables, giving rise to an effectively one-dimensional problem. The stationary Dirac equation is reduced to an effective one-dimensional Schr\"odinger equation with a fixed value of the  (conserved) longitudinal momentum. 
The models with translational symmetry in one direction, where either the magnetic field or the vector potential are piecewise constant, were discussed e.g. in \cite{DeMartino}, \cite{sharma}, \cite{Peeters}, \cite{ghosh}, and \cite{myoung}.  The exactly solvable systems with smooth magnetic field analyzed in the literature were usually related to the known, exactly solvable models of the nonrelativistic quantum mechanics \cite{Milpas}, \cite{negro}, \cite{TKGhosh},  or constructed from them with the use of supersymmetric techniques \cite{AstorgaNegro}, \cite{FernandezMidya}, \cite{Axel1}, \cite{Axel2}, \cite{isomorph}, \cite{correa}, \cite{BDG}.

In the current article,
we will also utilize translational symmetry to reduce the considered system into the effectively one-dimensional setting. The corresponding one-dimensional Schr\"odinger equation represents the new exactly solvable model presented here for the first time.  We discuss the spectral and scattering properties of both the effective one-dimensional and the initial two-dimensional relativistic system in dependence on the conserved longitudinal momentum and the strength of the coupling parameter.

The article is organized as follows: in the next section, we introduce the model with asymmetric step vector potential, we derive effectively one-dimensional stationary equation and find its explicit solutions in terms of Gauss hypergeometric functions. In the Section~\ref{Bound states}, discrete energies of the effective one-dimensional system are discussed whereas in the Section~\ref{Scattering}, we focus on the scattering properties. The last section is left for discussion.

\section{Solvable model of an asymmetric step potential}
Let us introduce the following Hamiltonian
\begin{equation}\label{h}
 H_D=-i\hbar\sigma_1\partial_x-i\hbar\sigma_2\partial_y+\frac{V_1}{\sqrt{1+e^{\frac{2x}{\sigma}}}}\sigma_2,
\end{equation}
where the coupling constant $V_1$ is fixed to be real and positive number\footnote{The spectrum and eigenstates of $H_D$ for negative values of $V_1$ can be obtained directly with the use of the operator $P_y\sigma_1$ where $P_yf(x,y)=f(x,-y)$.} and $\sigma_1$, $\sigma_2$ and $\sigma_3$ are Pauli matrices. We use the units where $\hbar=1$ and we fix $\sigma=1$.
The vector potential represents an asymmetric step. Indeed, defining the reflection operator as ${\cal{P}}=P\sigma_3$, where $PxP=-x$ and $PyP=-y$, the potential term breaks manifestly the reflection symmetry of $H_D$.
As (\ref{h}) commutes with the generator of translations $\hat p_y=-i\partial_y$, it possesses translational symmetry along $y$-axis.

Making the partial Fourier transform in the $y$ coordinate, we can rewrite $H_D$ in the form of direct integral
\begin{equation}\nonumber
H_D=\int^{\oplus}_{\mathbb{R}}H_D[k_y],	
\end{equation}
where the effectively one-dimensional Hamiltonian $H_D[k_y]$ is acting in $x$-coordinate with $k_y$ being fixed. Its explicit form reads
\begin{equation}\label{hef}
 H_D[k_y]=-i\sigma_1\partial_x+W(x)\sigma_2,
\end{equation}
where
\begin{equation}\label{W}
W(x)=k_y+\frac{V_1}{z(x)},\quad z(x)=\sqrt{1+e^{2x}}.
\end{equation}
We fix the domain of $H_D[k_y]$ as the Sobolev space $W^{1,2}(\mathbb{R},\mathbb{C}^2)$, the space of $\mathbb{C}^2$-valued square integrable functions with square integrable first derivatives. As the potential term $W(x)$ is smooth and bounded, it follows from the Kato-Rellich theorem that the operator $H_D[k_y]$ is self-adjoint on this domain \cite{RSII}. Once the spectrum of $H_D[k_y]$ is found, it is straightforward to reconstruct the spectrum $\sigma(H_D)$ of $H_D$. In our case, it can be understood as the union of $\sigma(H_D[k_y])$,  see \cite{RSIV} for more details.

It is convenient to inspect the stationary equation for the squared Dirac Hamiltonian $H_D[k_y]^2$. The operator reads explicitly
\begin{equation}\label{HD2}
 H_D[k_y]^2=\left(\begin{array}{cc}H^+_S[k_y]&0\\0&H^-_S[k_y]\end{array}\right),\quad H^\epsilon_S[k_y]=-\partial_x^2+W^2+\epsilon\,\partial_xW,\quad \epsilon=\pm.
\end{equation}
Let us focus on the spectral properties of $H_S[k_y]\equiv H^+_S[k_y]$
as those of $H_D[k_y]$ can be deduced in a straightforward manner afterwards. The explicit form of the associated nonrelativistic stationary equation reads
\begin{equation}\label{SE2}\left(-\partial_x^2+k_y^2+\frac{V_1(2k_y-1)}{z(x)}+\frac{V_1^2 }{z(x)^2}+\frac{V_1 }{z(x)^3}\right)\psi_{E,\pm}=E\psi_{E,\pm}.\end{equation}
It is the key finding of this article that this equation is exactly solvable for any $E$. To our best knowledge, the explicit solution of (\ref{SE2}) has not been discussed yet. The potential term in (\ref{SE2}) is generalization of the one discussed in \cite{artur1}.
With Eckart, P\"oschl-Teller and third hypergeometric \cite{artur1} potentials,  the potential term in (\ref{SE2}) is the new member of the family where the wave functions can be expressed in terms of Gauss hypergeometric functions. Contrary to the first two mentioned potentials, the eigenstates of the Hamiltonian $H_S[k_y]$ form an irreducible combination of two hypergeometric functions. To this instance, the potential belongs to the solvable potentials that have been explored recently \cite{artur1}, \cite{artur2}, \cite{artur3}, \cite{artur4}.

We prefer to leave the technical details of the solution of (\ref{SE2}) to the Appendix and present here the obtained eigenstates of $H_S[k_y]$. They can be written as
\begin{equation}\label{psi1}
 \psi_{E,\pm}=(z+1)^{\pm \alpha_1}(z-1)^{\alpha_2}u_\pm(z),
\end{equation}
where
\begin{eqnarray}\label{psi1u}
u_{\pm}(z)&=&{}_2F_1\left(\pm\alpha_1+\alpha_2-\alpha_3-1,\pm\alpha_1+\alpha_2+\alpha_3, \pm2\alpha_1,\frac{z+1}{2}\right)\nonumber\\&&+\frac{\left(-V_1\mp\alpha_1+\alpha_2+\left(\pm\alpha_1+\alpha_2+\alpha_3\right)z\right)}{\pm\,4\alpha_1}\times\nonumber\\&&{}_2F_1\left(\pm\alpha_1+\alpha_2-\alpha_3,\pm\alpha_1+\alpha_2+\alpha_3+1,1\pm2\alpha_1,\frac{z+1}{2}\right).
\end{eqnarray}
The quantities $\alpha_1$, $\alpha_2$ and $\alpha_3$ are defined as
\begin{align}\label{alphas}
\alpha_1=\frac{1}{2}\sqrt{-E+(k_y-V_1)^2},\quad
\alpha_2=\frac{1}{2}\sqrt{-E+(k_y+V_1)^2},\quad \alpha_3=\sqrt{-E+k_y^2}.
\end{align}
Let us notice that the argument $\frac{z+1}{2}$ in (\ref{psi1u}) falls out of the radius of convergence of the hypergeometric functions. In the Section~\ref{3.1}, where the asymptotic behavior of the wave function (\ref{psi1}) is discussed, the functions (\ref{psi1u})  will be transformed into the expressions with guaranteed convergence.

Let us consider the eigenstates $\Psi_{\sqrt{E},\pm}$ of the associated Dirac Hamiltonian $H_D[k_y]$. We have
\begin{equation}\label{Psipsi}
 \Psi_{\sqrt{E},\pm}=\left(\begin{array}{c}\psi_{E,\pm}\\\frac{-i\partial_x+iW}{\sqrt{E}}\psi_{E,\pm}\end{array}\right),\quad \Psi_{-\sqrt{E},\pm}=\sigma_3 \Psi_{\sqrt{E},\pm},
\end{equation}
and
\begin{equation}\label{eqzm}
 (H_D[k_y]\mp\sqrt{E})\Psi_{\pm\sqrt{E},\pm}=0.
\end{equation}
The spectrum of $H_D[k_y]$ is symmetric with respect to zero. This was to be expected as we deal with a chiral Hamiltonian; it anticommutes with $\sigma_3$, $\{\sigma_3,H_D[k_y]\}=0$.
The zero modes of $H_D[k_y]$ can be found by direct calculation as
\begin{equation}\label{zm0}
 \Psi_{0,-}=\left(\exp \left(-\int^{x}_0 W(s)ds\right),0\right)^T,\quad \Psi_{0,+}=\left(0,\exp \left(\int^{x}_0 W(s)ds\right),0\right)^T.
\end{equation}
Their square integrability depends on the asymptotic behavior of $W(x)$ that varies in dependence on $k_y$. We will discuss this point in Section \ref{3.2} in more detail.

\section{Bound states\label{Bound states}}
We focus on the non-relativistic system described by the Hamiltonian $H_S[k_y]$. We find its bound states by the asymptotic analysis of the wave functions (\ref{psi1}) in dependence on the coupling constant $V_1$ and on the values of $k_y$. As we mentioned above, the spectral properties of the effective one-dimensional operator $H_D[k_y]$ and the initial two-dimensional Dirac Hamiltonian $H_D$ can be deduced afterwards in straightforward manner.
\subsection{Asymptotics of the wave functions\label{3.1}}
As the potential term of $H_S[k_y]$ is smooth and asymptotically constant,
\begin{equation}\label{potasympt}
 W(x)\rightarrow \begin{cases}k_y,\quad x\rightarrow +\infty,\\ k_y+V_1,\quad x\rightarrow -\infty,\end{cases}\quad W^2(x)+\partial_xW(x)\rightarrow \begin{cases}k_y^2,\quad x\rightarrow \infty,\\ (k_y+V_1)^2,\quad x\rightarrow -\infty,\end{cases}
\end{equation}
the wave functions have to coincide asymptotically with a linear combination of the free particle solutions. In order to find the coefficients of the linear combination, we use the following formulas that turn the  hypergeometric functions into the form convergent at $x\rightarrow\pm\infty$:
\begin{eqnarray}
_2F_1(a,b,c,z)&=&\frac{\Gamma(c)\Gamma(b-a)}{\Gamma(b)\Gamma(c-a)}(-z)^{-a}{}_2F_1\left(a,1-c+a,1-b+a,\frac{1}{z}\right)\\&&+\frac{\Gamma(c)\Gamma(a-b)}{\Gamma(a)\Gamma(c-b)}(-z)^{-b}{}_2F_1\left(b,1-c+b,1-a+b,\frac{1}{z}\right),\\
{}_2F_1(a,b,c,z)&=&\frac{\Gamma(c)\Gamma(c-a-b)}{\Gamma(c-b)\Gamma(c-a)}z^{-a}{}_2F_1\left(a,1-c+a,1+b+a-c,1-\frac{1}{z}\right)\\&&+\frac{\Gamma(c)\Gamma(a+b-c)}{\Gamma(a)\Gamma(b)}(-1)^{c-a-b}(-1+z)^{c-a-b}z^{a-c}\nonumber\\&&\times{}_2F_1\left(c-a,1-a,1+c-a-b,1-\frac{1}{z}\right).
\end{eqnarray}
Then we find that the functions (\ref{psi1})
behave asymptotically as
\begin{eqnarray}\label{psiasym}
\psi_{E,\pm}&\sim& a^R_{\pm}e^{-\sqrt{E+k_y^2}}+b^R_{\pm}e^{\sqrt{E+k_y^2}},\ x\rightarrow +\infty,\\
\psi_{E,\pm}&\sim& a^L_{\pm}e^{-\sqrt{E+(k_y+V_1)^2}}+b^L_{\pm}e^{\sqrt{E+(k_y+V_1)^2}},\ x\rightarrow -\infty,
\end{eqnarray}
where the upper indices $L, R$ denote the asymptotics at $x\rightarrow\pm\infty$, and the lower indices $\pm$ distinguish the two independent solutions.
The coefficients can be written explicitly as
\begin{eqnarray}\label{coeffs}
a^R_{+}&=&\frac{(-\frac{1}{2})^{-\alpha_3-\alpha_1-\alpha_2}(1+2\alpha_3)\Gamma(-1-2\alpha_3)\Gamma(2
\alpha_1)}{\Gamma(\alpha_1-\alpha_2-\alpha_3)\Gamma(\alpha_1+\alpha_2-\alpha_3)},\nonumber\\
b^R_{+}&=&\frac{(-\frac{1}{2})^{\alpha_3-\alpha_1-\alpha_2}\Gamma(2
\alpha_1)\Gamma(2
\alpha_3)(\alpha_1^2-\alpha_2^2-\alpha_3V_1)}{\Gamma(1+\alpha_3+\alpha_1-\alpha_2)\Gamma(1+\alpha_3+\alpha_1+\alpha_2)},\nonumber\\
a^L_{+}&=&\frac{(-\frac{1}{2})^{-2\alpha_2}2^{-1+\alpha_1+\alpha_2}\Gamma(2\alpha_1)\Gamma(2
\alpha_2)(\alpha_3+2\alpha_2-V_1)}{\Gamma(\alpha_1+\alpha_2-\alpha_3)\Gamma(\alpha_1+\alpha_2+\alpha_3+1)},\nonumber\\
b^L_{+}&=&\frac{2^{-1+\alpha_1-\alpha_2}\Gamma(2\alpha_1)\Gamma(-2
\alpha_2)(\alpha_3-2\alpha_2-V_1)}{\Gamma(\alpha_1-\alpha_2-\alpha_3)\Gamma(\alpha_1-\alpha_2+\alpha_3+1)},\nonumber\\
a^R_{-}&=&a^R_{+}|_{\alpha_1\rightarrow-\alpha_1},\quad a^L_{-}=a^L_{+}|_{\alpha_1\rightarrow-\alpha_1}, \nonumber\\
b^R_{-}&=&b^R_{+}|_{\alpha_1\rightarrow-\alpha_1},\quad b^L_{-}=b^L_{+}|_{\alpha_1\rightarrow-\alpha_1}.
\end{eqnarray}

It is convenient to fix the combination that coincides with the single exponential function at $x\rightarrow +\infty$,
\begin{equation}\label{r1}
F\equiv\frac{b^R_{-}}{\lambda}\psi_{E,+}-\frac{b^R_{+}}{\lambda}\psi_{E,-}\sim e^{-\sqrt{k_y^2-E}x},\quad x\rightarrow +\infty.
\end{equation}
Here, we denoted $\lambda=\frac{(-\frac{1}{2})^{-2\alpha_2}(\alpha_2^2-\alpha_1^2+\alpha_3V_1)}{4\alpha_3\alpha_1}$.
At $x\rightarrow -\infty$, the function $F$ behaves in the following manner
\begin{equation}\label{r2}
 F\sim C_1e^{-\sqrt{(k_y+V_1)^2-E}x}+C_2e^{\sqrt{(k_y+V_1)^2-E}x},\quad x\rightarrow -\infty,
\end{equation}
where $C_1$ and $C_2$ are complex numbers given in terms of $\alpha_1$, $\alpha_2$, $\alpha_3$ and $V_1$.

The relations (\ref{r1}) and (\ref{r2}) suggest that the function $F$ will vanish exponentially for large $|x|$ provided that
\begin{equation}\label{cond1}
 \sqrt{k_y^2-E}>0,\quad \sqrt{(k_y+V_1)^2-E}>0,\quad C_1=0.
\end{equation}
The first two relations fix the range of allowed energies to the interval whose upper bound depends on the values of $k_y$,
\begin{equation}\label{Ebound}
 E\in[0,E_{max}),\quad E_{max}=\begin{cases}k_y^2,\quad k_y\geq-\frac{V_1}{2},\quad k_y\neq0, \\(k_y+V_1)^2, \quad k_y<-\frac{V_1}{2},\quad k_y\neq-V_1.\end{cases}
\end{equation}
Here, the lower bound follows from the fact that $E$ has to be non-negative as it should correspond to the eigenvalue of the operator that is related to the square of the self-adjoint Hamiltonian, see (\ref{HD2}) and (\ref{SE2}). When $k_y=0$ or $k_y=-V_1$, the first two conditions in (\ref{cond1}) cannot be fulfilled for $E\geq 0$.
To analyze the third condition, it is convenient to bring the coefficient $C_1$ into the following simplified form
\begin{equation}\label{C_1}
 C_1=\frac{2^{2\alpha_2-\alpha_3}\alpha_3\Gamma(2\alpha_3)\Gamma(2\alpha_2)(2\alpha_2+\alpha_3-V_1)}{\Gamma(1-\alpha_1+\alpha_2+\alpha_3)\Gamma(1+\alpha_1+\alpha_2+\alpha_3)}.
\end{equation}
The parameters $\alpha_1$, $\alpha_2$ and $\alpha_3$ are positive by definition (\ref{alphas}) and by the requirement (\ref{cond1}).
It implies that the terms $\Gamma(2\alpha_3)$ and $\Gamma(2\alpha_2)$ are nonvanishing.
Therefore, the coefficient $C_1$ can vanish as long as either $2\alpha_2+\alpha_3-V_1=0$ or $1-\alpha_1+\alpha_2+\alpha_3$ is equal to a non-positive integer. In the latter case, the $\Gamma$ function in the denominator would have singularity and would make the coefficient $C_1$ vanishing. Let us consider these cases in more detail.

\subsection{Zero modes\label{3.2}}
First, let us consider the case where $C_1$ vanishes due to $2\alpha_2+\alpha_3-V_1=0$. Substituting into the equation from (\ref{alphas}), we get 
\begin{equation}
 \alpha_3+2\alpha_2-V_1=\sqrt{k_y^2-E}-V_1+\sqrt{(k_y+V_1)^2-E}=0.
\end{equation}
The the expression on the left-hand side is monotonically decreasing function of $E$. Hence, the equation above has a solution if and only if the left-hand side has different signs at $E=0$ and $E=E_{\max}$, or it is identically zero at $E=0$, see (\ref{Ebound}). Straightforward analysis shows that this is the case for
\begin{equation}\label{zm}
 -V_1<k_y<0.
\end{equation}
It is consistent with the the discussion below (\ref{eqzm}); one can see from the explicit formulas for the zero modes (\ref{zm0})  that they are square integrable provided that the vector potential $W(x)$ changes sign asymptotically. Taking into account (\ref{potasympt}), we can see that this happens exactly for the values of $k_y$ specified in (\ref{zm}). The square integrable zero modes acquire the following explicit form where $k_y$ is supposed to satisfy (\ref{zm}),
\begin{equation}
 \Psi_0=(\psi_0,0)^T,\quad H_D[k_y]\Psi_0=0,\quad \psi_0=e^{k_yx}\left(\frac{-1+\sqrt{1+e^{2x}}}{1+\sqrt{1+e^{2x}}}\right)^{\frac{V_1}{2}},\quad H_S[k_y]\psi_0=0.
\end{equation}

\subsection{Existence of eigenstates with nonzero eigenvalues}
Now, let us focus on the cases where $C_1=0$ due to the singularity of the $\Gamma$ function in the denominator of (\ref{C_1}).  As $\alpha_1$, $\alpha_2$ and $\alpha_3$ are positive numbers, the only term that can be negative is
\begin{equation}
 I(E)\equiv\alpha_3-\alpha_1+\alpha_2=\sqrt{k_y^2-E}-\frac{1}{2}\sqrt{(k_y-V_1)^2-E}+\frac{1}{2}\sqrt{(k_y+V_1)^2-E}.
\end{equation}
The values of $E$ where $I(E)$ acquires negative integer values will correspond to the bound state energies of $H_S[k_y]$.
Let us inspect the behavior of $I(E)$ for different values of $k_y$. We will discuss the cases of positive and negative $k_y$ separately.

\subsubsection{$k_y\geq0$}
We will show that $I(E)$ is strictly positive by finding its positive lower bound. We have
\begin{equation}
 -\sqrt{(k_y-V_1)^2-E}\geq -\sqrt{k_y^2+V_1^2-E},\quad \sqrt{(k_y+V_1)^2-E}\geq \sqrt{k_y^2+V_1^2-E}.
\end{equation}
Then, taking into account (\ref{cond1}), we can write
\begin{equation}
 I(E)\geq \sqrt{k_y^2-E}-\frac{1}{2}\sqrt{k_y^2+V_1^2-E}+\frac{1}{2}\sqrt{k_y^2+V_1^2-E}=\sqrt{k_y^2-E}>0.
\end{equation}
It implies that $I(E)$ is always positive. Therefore, the coefficient $C_1$ does not vanish for any value of $E$. We conclude that the Hamiltonian $H_S[k_y]$, and, consequently, $H_D[k_y]$ have no bound states for any $k_y\geq0$.

\subsubsection{$k_y<0$}
We show that $I(E)$ is a decreasing function of $E$ for $k_y<0$. Indeed, its derivative,
\begin{equation}\label{dI}
 2\partial_E I(E)=-\frac{1}{\sqrt{k_y^2-E}}-\frac{1}{2\sqrt{(|k_y|-V_1)^2-E}}+\frac{1}{2 \sqrt{(|k_y|+V_1)^2-E}},
\end{equation}
can be bounded from above by a negative function. It can be obtained by using the following estimates in (\ref{dI}),
\begin{equation}
 \sqrt{(|k_y|+V_1)^2-E}\geq \sqrt{k_y^2+V_1^2-E},\quad \sqrt{(|k_y|-V_1)^2-E}\leq \sqrt{k_y^2+V_1^2-E}.
\end{equation}
It yields
\begin{equation}
 \partial_E I\leq -\frac{1}{\sqrt{k_y^2-E}}<0.
\end{equation}
Accordingly,  $I(E)$ is decreasing function of $E $ on the interval $[0,E_{max})$. It acquires its maximum at $E=0$,
\begin{equation}
 \max_{E\in[0,E_{\max})} I(E)=I(0)=\begin{cases} 0,\quad \mbox{for}\ \ -V_1<k_y<0,\\|k_y|-V_1,\quad  \mbox{for}\ k_y<-V_1,\end{cases}
\end{equation}
whereas its infimum is obtained at $E=E_{max}$, see (\ref{Ebound}),
\begin{equation}\label{minI}
 \inf_{E\in[0,E_{\max})} I(E)=I(E_{\max})=\begin{cases}\sqrt{2|k_y|V_1-V_1^2}-\sqrt{|k_y|V_1},\quad k_y\in(-\infty,-\frac{V_1}{2}],\\
\frac{V_1}{2}\left(\sqrt{1-2\frac{|k_y|}{V_1}}-\sqrt{1+2\frac{|k_y|}{V_1}}\right),\quad k_y\in(-\frac{V_1}{2},0).
\end{cases}
\end{equation}
When $k_y\leq-V_1$, we can write
\begin{eqnarray}
 I(E_{\max})&=&V_1\left(\sqrt{\frac{2|k_y|}{V_1}-1}-\sqrt{\frac{|k_y|}{V_1}}\right)\geq V_1\left(\sqrt{\frac{|k_y|}{V_1}+\frac{V_1}{V_1}-1}-\sqrt{\frac{|k_y|}{V_1}}\right)= 0.
\end{eqnarray}
Therefore, $I(E)$ is strictly positive and $H_S[k_y]$ has no bound states for $k_y\leq-V_1$.

For $k_y\in (-V_1,0)$, the number of bound states of $H_S[k_y]$ is given by the absolute value of the integer part of (\ref{minI}), $|[I(E_{\max})]| $, where we denoted the integer part of  by $[.]$. As the bound states of $H_D[k_y]$ can be obtained from those of $H_S[k_y]$ via (\ref{Psipsi}), we can conclude that the number of bound states $n$ of $H_D[k_y]$ as a function of $k_y$ can be expressed as
\begin{equation}\label{n}
 n= 2|[ I(E_{\max})]|+1.
\end{equation}
Here, the multiplicative factor $2$ reflects presence of negative energies while the summand $1$ stands for the single zero mode. As the function of $k_y$, $E_{\max}$ acquires its largest value for $k_y=-\frac{V_1}{2}$. Substituting this value to (\ref{minI}) and inserting the result into (\ref{n}), we get that  Hamiltonian $H_D[k_y]$ has maximum number of bound states $n_{max}=2\left[\frac{V_1}{\sqrt{2}}\right]+1$.

We solved the equation $I(E)=0$ numerically and reconstructed the spectrum of $H_S[k_y]$ and $H_D[k_y]$ in dependence on $k_y$, see Fig.~1. The probability densities of the bound states associated with the discrete energies of $H_S[k_y]$ and $H_D[k_y]$ are presented in Fig.~2.
\begin{figure}
\begin{center}
 \includegraphics{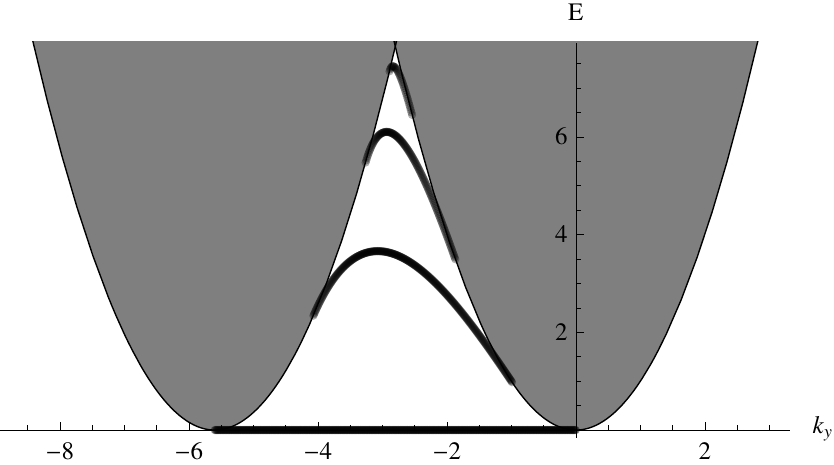} \includegraphics{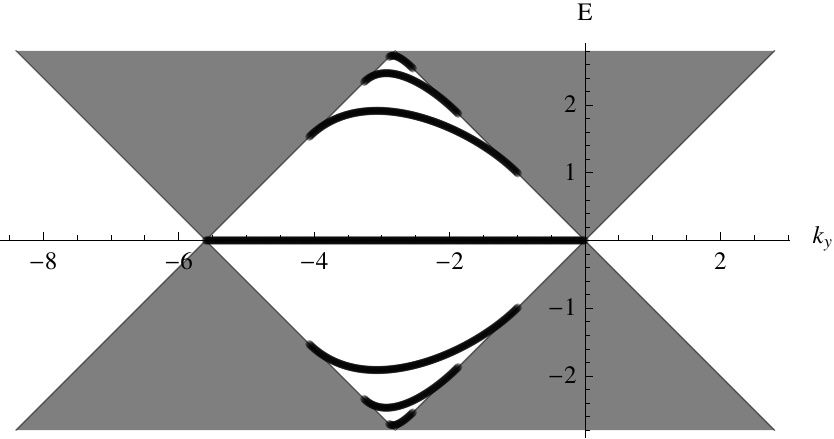}
\caption{Spectrum of the Schr\"odinger operator $H_S[k_y]$ (left) and of the Dirac operator $H_D[k_y]$ (right) in dependence on the value of $k_y$. We fixed $V_1=5.6$. In accordance with the formula for the maximum number of bound states $n_{max}=2\left[\frac{V_1}{\sqrt{2}}\right]+1$, the Hamiltonian $H_D[k_y]$ has seven bound states for $k_y=-\frac{V_1}{2}$.}
\end{center}
\end{figure}

\begin{center}
\begin{figure}
\centering
	\begin{tabular}{cc}
 \includegraphics[scale=.7]{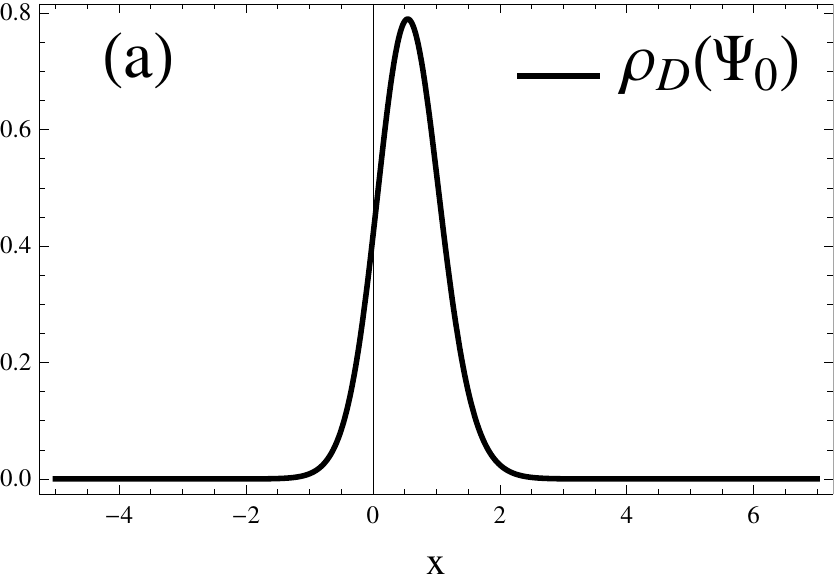}\hspace{10mm}&\includegraphics[scale=.7]{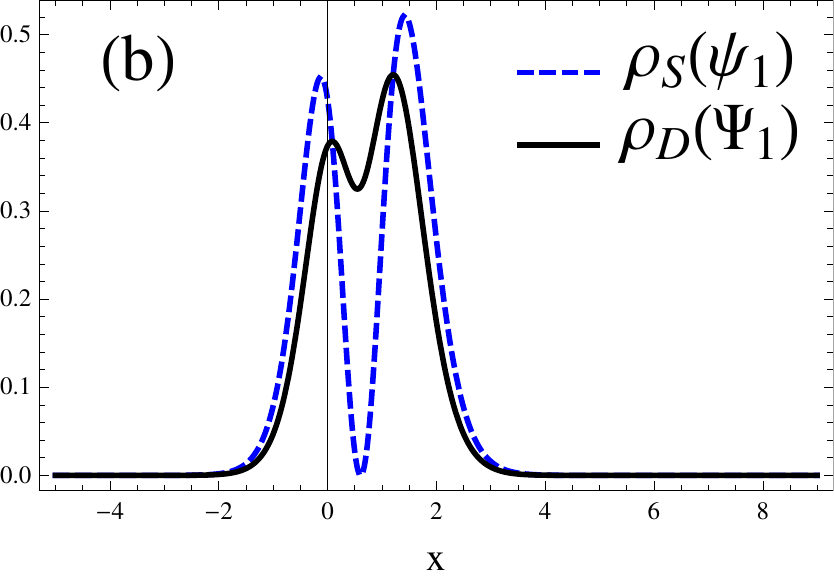}\\\includegraphics[scale=.7]{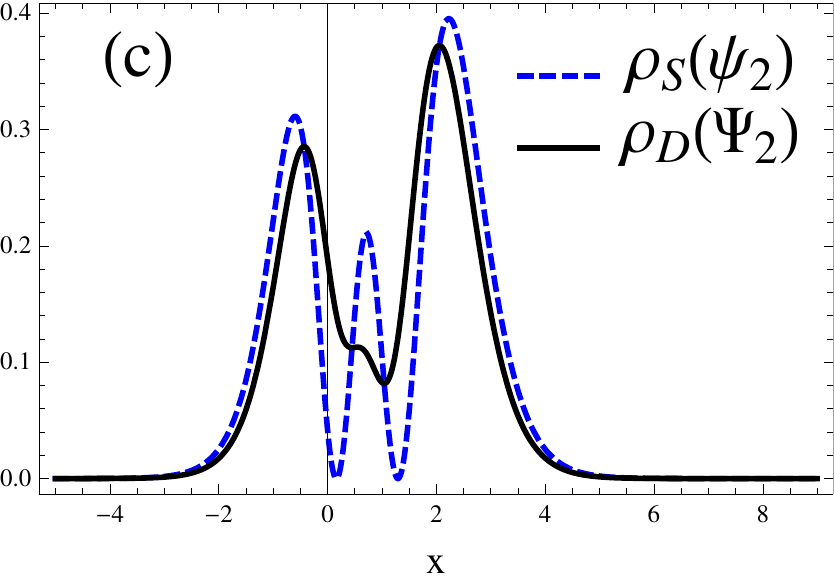}&\includegraphics[scale=.7]{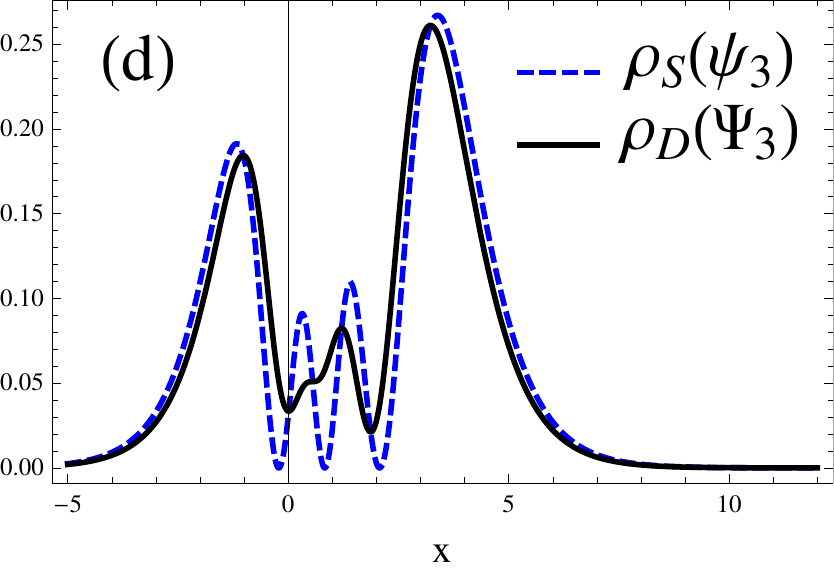}
\end{tabular}\caption{Probability densities of (normalized) bound states $\psi_j$ and $\Psi_j$, $j=0,1,2,3$, of the effective Schr\"odinger and Dirac Hamiltonians $H_S[k_y]$ and $H_D[k_y]$, respectively. Here we used $\rho_S(f)=\overline{f}f$ and $\rho_D(F)=F^{\dagger}F$. We fixed $V_1=5.6$ and $k_y=-\frac{V_1}{2}$. We did not plot probability densities for eigenstates of negative energies as there holds $\rho_D(\sigma_3\Psi_j)=\Psi_3$. }
\end{figure}
\end{center}

\section{Scattering\label{Scattering}}
Scattering properties of $H_D[k_y]$ can be unambiguously determined from those of $H_S[k_y]$ with the use of (\ref{Psipsi}). Henceforth, let us analyze the reflection and transmission coefficients of the latter Hamiltonian.

First, we consider the wave function propagating from the left, i.e. we require it to have the following asymptotic form,
\begin{equation}\label{psil}
 \psi_l=\begin{cases}e^{x\sqrt{(k_y+V_1)^2-E}}+\mathcal{R}_le^{-x\sqrt{(k_y+V_1)^2-E}},\quad x\rightarrow-\infty,\\
       \tilde{\mathcal{T}}_le^{x\sqrt{k_y^2-E}},\quad x\rightarrow\infty.
\end{cases}
\end{equation}
The incoming wave exists for $E>(k_y+V_1)^2$, and the outgoing (scattering) wave for $E>k_y^2$. When $(k_y+V_1)^2<k_y^2$ and $E\in((k_y+V_1)^2,k_y^2)$, then the transmitted wave is exponentially vanishing and total reflection takes place. Using (\ref{psiasym}), we can find the coefficients $\mathcal{R}_l$ and $\tilde{\mathcal{T}}_l$ in terms of (\ref{coeffs}),
\begin{equation}
 \mathcal{R}_l=\frac{a^R_{-}a^L_{+}-a^R_{+}a^L_{-}}{a^R_{-}b^L_{+}-a^R_{+}b^L_{-}},\quad \tilde{\mathcal{T}}_l=\frac{a^R_{-}b^R_{+}-a^R_{+}b^R_{-}}{a^R_{-}b^L_{+}-a^R_{+}b^L_{-}}.
\end{equation}
The quantity $\mathcal{R}_l$ coincides with the reflection coefficient. However, $\tilde{\mathcal{T}}_l$ is proportional but not equal to the transmission coefficient in general.
It is caused by the fact that the potential converges to different values at $x\rightarrow\pm\infty$, see e.g. \cite{Griffits}.
 With the use of the continuity equation, we get the following relation for the transmission coefficient $\mathcal{T}_l$,
\begin{eqnarray}\label{TRl}
&&|\mathcal{R}_l|^2+|\mathcal{T}_l|^2=1 \quad\mbox{for}\quad \mathcal{T}_l=\sqrt{\frac{\sqrt{E-k_y^2}}{\sqrt{E-(k_y+V_1)^2}}}\,\tilde{\mathcal{T}}_l.
\end{eqnarray}

Now, let us consider the wave function traveling from the right. It behaves asymptotically as
\begin{equation}\label{psir}
 \psi_{r}=\begin{cases}e^{-x\sqrt{k_y^2-E}}+\mathcal{R}_r e^{x\sqrt{k_y^2-E}},\quad x\rightarrow\infty,\\
       \tilde{\mathcal{T}}_re^{-x\sqrt{(k_y+V_1)^2-E}},\quad x\rightarrow-\infty.
\end{cases}
\end{equation}
This time, the incoming (scattering) wave exists for $E>k_y^2$ while the outgoing wave for $E>(k_y+V_1)^2$. When $(k_y+V_1)^2>k_y^2$ and $E\in(k_y^2,(k_y+V_1)^2)$, there are no solutions for the outgoing wave in the form of scattering states.
We can find the coefficients as
$$\mathcal{R}_r=\frac{b^L_{-}b^R_{+}-b^L_{+}b^R_{-}}{b^L_{-}a^R_{+}-b^L_{+}a^R_{-}},\quad\tilde{\mathcal{T}}_r=\frac{b^L_{-}a^L_{+}-b^L_{+}a^L_{-}}{b^L_{-}a^R_{+}-b^L_{+}a^R_{-}}.  $$
Like in the previous case, $\mathcal{R}_r$ represents the reflection coefficient. We can find transmission coefficient $\mathcal{T}_r$ such that
\begin{eqnarray}\label{TRr}
 &&|\mathcal{R}_r|^2+|\mathcal{T}_r|^2=1 \quad\mbox{for}\quad \mathcal{T}_r=\sqrt{\frac{\sqrt{E-(k_y+V_1)^2}}{\sqrt{E-k_y^2}}}\tilde{\mathcal{T}}_r.
\end{eqnarray}

\begin{figure}
\begin{center}
 \includegraphics{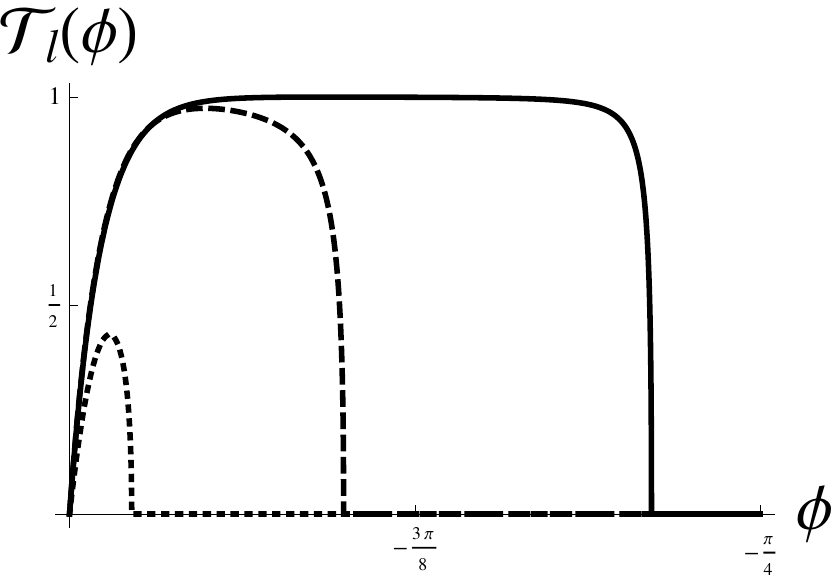} \label{T}
\caption{Transmission amplitude $\mathcal{T}_l$ for fixed energy $E$ and $k_y=\sqrt{E}\sin\phi$, $\phi\in(-\pi/2,\pi/2)$. We fixed $V_1=4$ and $E=4.01$ (dotted black curve),  $E=4.2$ (dashed black) and $E=5$ (solid black curve).}
\end{center}
\end{figure}

The formulas (\ref{TRl}) and (\ref{TRr}) for $\mathcal{T}_l$ and $\mathcal{T}_r$ imply that the transmission occurs only for $E$ and $k_y$ satisfying $E-k_y^2>0$ and $E-(k_y+V_1)^2>0$. Fixing the energy of the incoming wave, the two inequalities hold for a limited range of $k_y$,
\begin{equation}\label{int}
 k_y\in(-\sqrt{E},-V_1+\sqrt{E}),\quad \sqrt{E}>\frac{V_1}{2},
\end{equation}
where the restriction of $\sqrt{E}$ on the right-hand side ensures that the interval is non-empty. When $\sqrt{E}\leq\frac{V_1}{2}$ or $k_y$ does not fall into the specified interval (\ref{int}), then
the incident waves are totally reflected; the barrier is filtering the wave functions dependently on their energy and the value of the longitudinal momentum $k_y$. It also implies that there are angles behind the barrier in which the wave function cannot be scattered. Consider  the transmitted wave function $e^{ik_yy}\psi_l(x)$ that propagates to the right, see (\ref{psil}). We get $E=k_x^2+k_y^2$, where $k_x=\sqrt{E-k_y^2}$. We can parametrize $k_y=\sqrt{E}\sin\phi$, where $\phi\in(-\pi/2,\pi/2)$ denotes the angle in which the transmitted wave propagates away from the barrier. Substituting into (\ref{int}), we get that the transmission is possible only for $\phi\in \left(-\pi/2,\arcsin\left(1-\frac{V_1}{\sqrt{E}}\right)\right)
$, see Fig.~\ref{T}.
Let us mention that the tunneling dependent on the wave-vector through magnetic barriers was discussed for the non-relativistic particles e.g. in \cite{Matulis} and for the effectively one-dimensional Dirac system with symmetric vector potential in \cite{Milpas}.

\section{Discussion}

In the current article, we have presented the new exactly solvable model of two-dimensional Dirac fermions in the presence of an asymptotically vanishing, asymmetric, magnetic barrier (\ref{h}). The translation symmetry allowed us to work with effectively one-dimensional stationary equation, whose solutions were given in terms of an irreducible combination of two hypergeometric functions, see (\ref{psi1}) and (\ref{psi1u}).

The solution was facilitated by the fact that the Schr\"odinger equation for the upper component of the spinor can be reduced to the Heun equation and solved analytically. We elaborated its solution in detail in the Appendix. In this manner, the model extends the family of exactly solvable systems discovered recently, \cite{artur1}, \cite{artur2}, \cite{artur3}, \cite{artur4}.
It is also worth noticing in this context that there were analyzed quasi-exactly solvable system whose solutions were found to be written in terms of the confluent Heun functions \cite{Portnoi}.

The exactly solvable Schr\"odinger equation (\ref{SE2}) is generalization of the one discussed in \cite{artur1} whose potential term was $V_0+\frac{V_1}{\sqrt{1+\exp{2x}}}$. Careful comparison reveals one striking feature whose deeper meaning is yet to be understood; the latter potential of the \textit{Schr\"odinger} operator coincides with the vector potential of our \textit{Dirac} Hamiltonian, see (\ref{W}) here and (1) in \cite{artur1}.

We focused on the spectral properties of the two-dimensional Dirac Hamiltonian (\ref{h}). Enjoying the translation invariance of the system, its spectrum could be obtained from the analysis of the effective one-dimensional operator (\ref{hef}).
We found that the energies of the two-dimensional Dirac fermion strongly depend on the value of the longitudinal momentum $k_y$.   When $k_y$ is positive, the effective one-dimensional Hamiltonian has no bound states, whereas it can have finite number of bound states for negative values of $k_y$. The discrete energy levels of the one-dimensional Hamiltonian form spectral bands of the two-dimensional energy operator, see Fig.~1. As it was discussed in \cite{Dispersionless}, there can exist wave packets in the system that are dispersionless in perpendicular direction on the barrier.

Our system is similar to the P\"osch-Teller model with the (vector) potential $V=\tanh x$. However, the systems differ in their symmetries; the Hamiltonian of the P\"oschl-Teller system commutes with the reflection operator $\mathcal{P}=\sigma_3 P$, while our system breaks this symmetry explicitly. The difference is mani\-fested when the spectra of the two systems are compared. Energy of the Dirac particle in presence of P\"oschl-Teller potential is even function of $k_y$, whereas the spectrum of our system is asymmetric, compare Fig.~3 in \cite{Milpas} and Fig.~1 here.

Supersymmetric techniques can be used to generate the new exactly solvable systems from the one presented here. The (confluent) Darboux-Crum transformation was used for construction of the new solvable models e.g. from the harmonic oscillator, Rosen-Morse model, Coulomb potential or the free particle \cite{FernandezMidya}, \cite{Axel1}, \cite{correa}. The new systems obtained in this manner from either $H_S[k_y]$ or $H_D[k_y]$ would have the same spectrum as up to a finite number of discrete energy levels, provided that regularity of the new potential is guaranteed. As much as the application of the SUSY transformation on our model represents interesting research direction, we find it going beyond the scope of the present paper.

\section*{Acknowledgements}
The research by Artur Ishkhanyan has been supported by the State Committee of Science of the Republic of Armenia (project 18RF-139), the Armenian National Science and Education Fund (ANSEF grant PS-4986), the Russian-Armenian (Slavonic) University at the expense of the Ministry of Education and Science of the Russian Federation, as well as by the project "Leading research universities of Russia" (grant FTI\textunderscore 24 \textunderscore2016 of the Tomsk Polytechnic University). V\'i{}t Jakubsk\'y was supported by GA\v CR grant no. 15-07674Y.

\section*{Appendix. Solution of the associated Schr\"odinger equation (\ref{SE2}) }

\noindent

 The transformation of the dependent and independent variables
\begin{equation} \label{GrindEQ__1_}
\psi =\varphi \, (z)\; u(z),   z=z(x)
\end{equation}
reduces the one-dimensional Schrödinger equation
\begin{equation} \label{GrindEQ__2_}
\frac{d^{2} \psi }{dx^{2} } +\frac{2m}{\hbar ^{2} } \left(E-V(x)\right)\psi =0
\end{equation}
to the equation
\begin{equation} \label{GrindEQ__3_}
u_{zz} +\left(2\frac{\varphi _{z} }{\varphi } +\frac{\rho _{z} }{\rho } \right)\; u_{z} +\; \left(\frac{\varphi _{zz} }{\varphi } +\frac{\rho _{z} }{\rho } \frac{\varphi _{z} }{\varphi } +\frac{2m}{\hbar ^{2} } \frac{E-V(z)}{\rho ^{2} } \right)\; u=0,
\end{equation}
where $\rho =dz/dx$.

 Consider a potential given as
\begin{equation} \label{GrindEQ__4_}
V(z)=V_{0} +\frac{V_{1} }{z} +\frac{V_{2} }{z^{2} } +\frac{V_{3} }{z^{3} }
\end{equation}
with
\begin{equation} \label{GrindEQ__5_}
z=\sqrt{1+e^{2x/\sigma } } .
\end{equation}
For this coordinate transformation we have
\begin{equation} \label{GrindEQ__6_}
\rho =\frac{dz}{dx} =\frac{(z+1)(z-1)}{\sigma z} .
\end{equation}
It is then readily checked by direct substitution that putting
\begin{equation} \label{GrindEQ__7_}
\varphi =(z+1)^{\alpha _{1} } (z-1)^{\alpha _{2} }
\end{equation}
reduces equation (\ref{GrindEQ__2_}) to the general Heun equation
\begin{equation} \label{GrindEQ__8_}
\frac{d^{2} u}{dz^{2} } +\left(\frac{\gamma }{z} +\frac{\delta }{z-1} +\frac{\varepsilon }{z+1} \right)\, \frac{du}{dz} +\frac{\alpha \
	 \beta \, z-q}{z(z-1)(z+1)} u=0,
\end{equation}
where the involved parameters are given as
\begin{equation} \label{GrindEQ__9_}
\left(\gamma ,\delta ,\varepsilon \right)=\left(-1,1+2\alpha _{2} ,1+2\alpha _{1} \right),
\end{equation}
\begin{equation} \label{GrindEQ__10_}
\alpha \beta =(\alpha _{1} +\alpha _{2} )^{2} +\frac{2m\sigma ^{2} }{\hbar ^{2} } \left(E{\kern 1pt} -V_{0} \right),
\end{equation}
\begin{equation} \label{GrindEQ__11_}
q=-\alpha _{1} +\alpha _{2} -\frac{2m\sigma ^{2} }{\hbar ^{2} } V_{3} \, .
\end{equation}
\begin{equation} \label{GrindEQ__12_}
\alpha _{1} =\pm \sqrt{\frac{m\sigma ^{2} }{2\hbar ^{2} } \left(-E+V_{0} -V_{1} +V_{2} -V_{3} \right)} ,
\end{equation}
\begin{equation} \label{GrindEQ__13_}
\alpha _{2} =\pm \sqrt{\frac{m\sigma ^{2} }{2\hbar ^{2} } \left(-E+V_{0} +V_{1} +V_{2} +V_{3} \right)} .
\end{equation}

 Now we attempt to solve the Heun equation (\ref{GrindEQ__8_}) in terms of the hypergeometric functions.

 A helpful observation here is that $\gamma =-1$. This means that the characteristic exponents $(0,1-\gamma )=(0,2)$ of the singularity $z=0$ differ by an integer, hence, the Frobenius solution of the Heun equation generally involves a logarithmic term. However, there is a case when the singularity becomes \textit{apparent} (simple), that is, when the logarithmic term disappears and the solution becomes analytic in the singularity. The condition for this is known to be (see, e.g., \cite{artur5}, \cite{Maier})
\begin{equation} \label{GrindEQ__14_}
q^{2} +q\left(\varepsilon -1+a(\delta -1)\right)+a\alpha \beta =0,
\end{equation}
where $a$ is the location of the third regular singularity of the Heun equation ($a=-1$ for equation (\ref{GrindEQ__8_})). This condition is readily derived by expanding the solution of the Heun equation as a Taylor series $u=\sum _{n=0}^{\infty }c_{n} z^{n}  $ with $c_{0} \ne 0$, and successively calculating the coefficients $c_{n} $ after substitution of the series into the Heun equation. In calculating $c_{2} $ a division by zero will occur, unless the accessory parameter $q$ satisfies equation (\ref{GrindEQ__14_}), in which case the equation for $c_{2} $ is identically satisfied.

 It is known that for a root of equation (\ref{GrindEQ__14_}) the Heun equation is solved in terms of the Gauss hypergeometric functions as \cite{artur5}
\begin{equation} \label{GrindEQ__15_}
u={}_{2} F_{1} \left(\alpha ,\beta ;\varepsilon -1;\frac{a-z}{a-1} \right)+\frac{q+a\left(\delta -1\right)}{\varepsilon -1} \cdot {}_{2} F_{1} \left(\alpha ,\beta ;\varepsilon ;\frac{a-z}{a-1} \right),
\end{equation}
With $\alpha ,\beta $ determined from (\ref{GrindEQ__10_}) together with the Fuchsian condition $1+\alpha +\beta =\gamma +\delta +\varepsilon $:
\begin{equation} \label{GrindEQ__16_}
\alpha ,\beta =\alpha _{1} +\alpha _{2} \pm \sqrt{\frac{2m\sigma ^{2} }{\hbar ^{2} } \left(-E+V_{0} \right)} ,
\end{equation}
and $a=-1$, this gives the solution of the Schrödinger equation
\begin{equation} \label{GrindEQ__17_}
\psi =(z+1)^{\alpha _{1} } (z-1)^{\alpha _{2} } \left({}_{2} F_{1} \left(\alpha ,\beta ;2\alpha _{1} ;\frac{z+1}{2} \right)+\frac{q-2\alpha _{2} }{2\alpha _{1} } \cdot {}_{2} F_{1} \left(\alpha ,\beta ;1+2\alpha _{1} ;\frac{z+1}{2} \right)\right).
\end{equation}

 The substitution of the parameters (\ref{GrindEQ__9_})-(\ref{GrindEQ__13_}) into equation (\ref{GrindEQ__14_}) reveals that the condition is satisfied if
\begin{equation} \label{GrindEQ__18_}
V_{2} =\frac{2m\sigma ^{2} V_{3}^{2} }{\hbar ^{2} } .
\end{equation}
With this, the potential (\ref{GrindEQ__4_}) becomes
\begin{equation} \label{GrindEQ__19_}
V(z)=V_{0} +\frac{V_{1} }{z} +\frac{V_{3} \left(1+2m\sigma ^{2} V_{3} z/\hbar ^{2} \right)}{z^{3} } .
\end{equation}
Putting now $2m=\hbar =\sigma =1$ and $\alpha _{3} =\sqrt{-E+k_{y}^{2} } $, and changing
\begin{equation} \label{GrindEQ__20_}
V_{0} \to k_{y}^{2} ,   V_{1} \to V_{1} \left(2k_{y} -1\right),   V_{3} \to V_{1} ,
\end{equation}
equation (\ref{GrindEQ__19_}) recovers the potential discussed in the present paper and equation (\ref{GrindEQ__17_}) produces the corresponding solution (note that the solution is valid for arbitrary combination of signs of $\alpha _{1} $ and $\alpha _{2} $).

\end{document}